\documentclass[12pt]{article}
\usepackage{a4wide}
\usepackage{epsfig}
\usepackage{amsmath}

\newlength{\absize}
\catcode`@=11
\def\citer{\@ifnextchar [{\@tempswatrue\@citexr}{\@tempswafalse\@citexr[]}}

\def\@citexr[#1]#2{\if@filesw\immediate
  \write\@auxout{\string\citation{#2}}\fi
  \def\@citea{}\@cite{\@for\@citeb:=#2\do
    {\@citea\def\@citea{--\penalty\@m}\@ifundefined
       {b@\@citeb}{{\bf ?}\@warning
       {Citation `\@citeb' on page \thepage \space undefined}}%
\hbox{\csname b@\@citeb\endcsname}}}{#1}}
\catcode`@=12

\begin{document}
  \thispagestyle{empty}
  \pagestyle{empty}
  \renewcommand{\thefootnote}{\fnsymbol{footnote}}
\newpage\normalsize
    \pagestyle{plain}
    \setlength{\baselineskip}{4ex}\par
    \setcounter{footnote}{0}
    \renewcommand{\thefootnote}{\arabic{footnote}}
\newcommand{\preprint}[1]{%
  \begin{flushright}
    \setlength{\baselineskip}{3ex} #1
  \end{flushright}}
\renewcommand{\title}[1]{%
  \begin{center}
    \LARGE #1
  \end{center}\par}
\renewcommand{\author}[1]{%
  \vspace{2ex}
  {\Large
   \begin{center}
     \setlength{\baselineskip}{3ex} #1 \par
   \end{center}}}
\renewcommand{\thanks}[1]{\footnote{#1}}
\vskip 0.5cm

\begin{center}
{\large \bf Induced Fractional Zero-Point Canonical Angular
Momentum on Charged Particles of Aharonov - Bohm Vector Potential
and ``Spectator" Magnetic Field}
\end{center}
\vspace{1cm}
\begin{center}
Jian-Zu Zhang
\end{center}
\begin{center}
Institute for Theoretical Physics, East China University of
Science and Technology, Box 316, Shanghai 200237, P. R. China
\end{center}
\vspace{1cm}

\begin{abstract}
The induced fractional zero-point canonical angular momentum on
charged particles by the Aharonov - Bohm (AB) vector potential is
realized via modified combined traps.  It explores new features
for this type of quantum effects: In a limit of vanishing
mechanical kinetic energy the AB vector potential alone cannot
induce a fractional zero-point canonical angular momentum on
charged particles at the quantum mechanical level in the AB
magnetic field-free region; But for the case of the AB vector
potential with another one of a ``spectator" magnetic field the AB
vector potential induces a fractional zero-point canonical angular
momentum in the same limit. The ``spectator" one does not
contribute to such a fractional zero-point quantity, but plays
essential role in guaranteeing non-trivial dynamics survived in
this limit at the quantum mechanical level. These results are
significance in investigations of the AB effects and related
fields for both theories and experiments.
\end{abstract}

\begin{flushleft}
\end{flushleft}
\clearpage
Ever since it has recognized that quantum states of charged
particles can be influenced by electromagnetic effects even
charged particles cannot access to the region of non-vanishing
fields \cite{ES,AB}, the Aharonov - Bohm (AB) effect \cite{AB}, as
a unusual quantum phase effect, has been received much attention
for years \cite{OP,PT89,M-K}. According to this the interference
spectrum should suffer a shift according to the amount of magnetic
flux enclosed by two electron beams, even through magnetic
field-free region. The AB effect is purely quantum mechanical
which explored far-reaching consequences of vector potential in
quantum theory. The effect has been confirmed experimentally
\cite{expt}. Since the discovering of this effect investigations
in this topic concentrated on revealing new types of quantum
phases: The Aharonov-Casher effect \cite{AC}, the
He-McKellar-Wilkens phase \cite{HMW} and the Anandan phase
\cite{Anan}.

In another aspect the induced angular momentum of the AB vector
potential has been predicted by Peshkin, Talmi and Tassie for over
forty years
\cite{PT89,PT60,TP,PTT,Pesh81,Pesh82}. This type of quantum
effects is so remarkable that in quantum mechanics the vector
potential itself has physical significant meaning and becomes
effectively measurable not only in shifts of interference spectra
originated from quantum phases but also in physical observables.

Recently Kastrup \cite{Kast06} considered the question of how to
quantize a classical system of the canonically conjugate pair
angle and orbital angular momentum which has been a controversial
issue since the founding days of quantum mechanics \cite{Kast03}.
The problem is that the angle is a multivalued or discontinuous
variable on the corresponding phase space. A crucial point is that
the irreducible unitary representations of the
euclidean group $E(2)$ or of its covering groups allow for orbital
angular momentum $l = \hbar(n+\delta)$ where $n=0,1,2,\cdots,$ and
$0\le \delta<1$. The case $\delta\ne 0$ corresponds to fractional
zero-point angular momentum.
Kastrup investigated the physical possibility of fractional
orbital angular momentum in connection with the quantum optics of
Laguerre-Gaussian laser modes in external magnetic fields, and
pointed out that if implementable this would lead to a wealth of
new theoretical, experimental and even technological
possibilities.

In this paper the induced fractional zero-point canonical angular
momentum on charged particles by the AB vector potential is
realized via modified combined traps.  It explores new features
for this type of quantum effects: In a limit of vanishing
mechanical kinetic energy the AB vector potential alone cannot
induce a fractional zero-point canonical angular momentum on
charged particles at the quantum mechanical level in the AB
magnetic field-free region; But for the case of the AB vector
potential with another one of a ``spectator" magnetic field the AB
vector potential induces a fractional zero-point canonical angular
momentum in the same limit. The ``spectator" one does not
contribute to such a fractional zero-point quantity, but plays
essential role in guaranteeing non-trivial dynamics survived in
this limit at the quantum mechanical level. These results are
significance in investigations of the AB effects and related
fields for both theories and experiments.

We consider ions constrained in a modified combined trap including
the AB type magnetic field. The Paul, Penning, and combined traps
share the same electrode structure \cite{DHST}. A combined trap
operates in all of the fields of the Paul and Penning traps being
applied simultaneously. The trapping mechanism in a Paul trap
involves an oscillating axially symmetric electric potential
$\tilde{U}(\rho,\phi,z,t)= U(\rho,\phi,z)cos\tilde{\Omega} t$
with
$U(\rho,\phi,z)=Vd^2(z^2-\rho^2/2)/2$
where $\rho$, $\phi$ and $z$ are cylindrical coordinates, $V$ and
$d$ are, respectively, characteristic voltage and length, and
$\tilde{\Omega}$ is a large radio-frequency. The dominant effect
of the oscillating potential is to add an oscillating phase factor
to the wave function. Rapidly varying terms of time in
Schro¡§dinger equation can be replaced by their average values.
Thus for
$\tilde{\Omega}\gg \Omega\equiv \left(\sqrt{2}|q V|\mu d^2
\right)^{1/2}$
we obtain a time-independent effective electric potential
$V_{eff}=q^2\nabla U\cdot\nabla U/4\mu\tilde{\Omega}^2=
\mu\omega_P^2(\rho^2+4z^2)/2$
where $\mu$ and $q$ are, respectively, the mass and charge of the
trapped particle or ion, and
$\omega_P=\Omega^2/4\tilde{\Omega}$ \cite{CSW}.
A modified combined trap combines the above electrostatic
potential and two magnetic fields: a homogeneous magnetic field
${\bf B}_c$ aligned along the $z$ axis in a normal combined trap
and an AB type magnetic field ${\bf B}_0$ produced by, for
example, an infinitely long solenoid with radius
$\rho=(x_1^2+x_2^2)^{1/2}=a$. Inside the solenoid $(\rho<a)\;$
${\bf B}_{0,in}=(0, 0, B_0)$ is homogeneous along the $z$ axis,
and outside the solenoid $(\rho>a)$ ${\bf B}_{0,out}=0$. The
vector potential $A_{c,i}$ of ${\bf B}_c$ is chosen as (Henceforth
the summation convention is used)
$A_{c,i}=-B_c\epsilon_{ij}x_j/2$, $A_{c,z}=0,\; (i,j=1,2).$
The AB vector potential ${\bf A}_0$ is: Inside the solenoid
$A_{0,i}=A_{in,i}=-B_0\epsilon_{ij}x_j/2$,\; $A_{in,z}=0;$
Outside the solenoid
$A_{0,i}=A_{out,i}=-B_0 a^2 \epsilon_{ij}x_j/2x_k
x_k,\;A_{out,z}=0,\; (i,j,k=1,2).$
At $\rho=a$ the potential ${\bf A}_{in}$ passes continuously over
into ${\bf A}_{out}$. The ion's motion is confined to be planar,
rotationally symmetric.
The Hamiltonian of the modified combined trap is
$H=\left(p_i- \frac{q}{c}A_{c,i}-\frac{q}{c}A_{0,i}\right)^2/2\mu
+p_z^2/2\mu+ \mu\omega_P^2\left(x_i^2+4z^2\right)/2$.
This Hamiltonian can be decomposed into a one-dimensional harmonic
Hamiltonian $H_z(z)$ along the $z$-axis with the axial frequency
$\omega_z=2\omega_P$ and a two-dimensional Hamiltonian. Inside the
solenoid the ion's motion is the same as the one with a total
magnetic field ${\bf B}_c+{\bf B}_{0,in}$. In the following we
consider the motion out the solenoid.

Outside the solenoid the two-dimensional Hamiltonian is
\cite{DHST,CSW}
\begin{equation}
\label{Eq:H2}
H_{\perp}(x_1,x_2)=\frac{1}{2\mu}\left(p_i+\frac{1}{2}\mu\omega_c
\epsilon_{ij}x_j+\frac{1}{2}\mu\omega_0
a^2\frac{\epsilon_{ij}x_j}{x_k x_k}\right)^2+
\frac{1}{2}\mu\omega_P^2 x_i^2,
\end{equation}
where $\omega_c=qB_c/\mu c$ and $\omega_0=qB_0/\mu c$ are the
cyclotron frequencies corresponding to, respectively, the magnetic
fields ${\bf B} _c$ and ${\bf B}_{0,in}$. The Hamiltonian
$H_{\perp}$ possess a rotational symmetry in $(x_1, x_2)$ - plane.
The $z$-component of the canonical orbital angular momentum
$J_z=\epsilon_{ij} x_i p_j$
commutes with $H_{\perp}$. They have common eigenstates.

We consider the limiting case of vanishing mechanical kinetic
energy $E_k=\mu \dot{x_i} \dot{x_i}/2$. In a laser trapping field,
using a number of laser beans and exploiting Zeeman tuning,
the speed of atoms can be slowed to the extent of $1\;ms^{-1}$,
see \cite{note-2}.
It should be emphasized that in order to experimentally realizing
the limit of vanishing mechanical kinetic energy through laser
cooling we should consider ions with inner structures, not point
particles. In the limiting case of vanishing mechanical kinetic
energy the Hamiltonian $H_{\perp}$ in Eq.~(\ref{Eq:H2}) has
non-trivial dynamics \cite{Baxt,JZZ96}. The Lagrangian
corresponding to $H_{\perp}$ is
$L=\mu\dot{x_i}\dot{x_i}/2 -\mu\omega_c
\epsilon_{ij}\dot{x_i}x_j/2-\mu\omega_0
a^2\epsilon_{ij}\dot{x_i}x_j/2x_k x_k -\mu\omega_P^2 x_i x_i/2.$
In the limit of vanishing mechanical kinetic energy in the
Hamiltonian formalism,\\
$E_k=\mu \dot{x_i} \dot{x_i}/2=\left(p_i
+\mu\omega_c \epsilon_{ij}x_j/2+\mu\omega_0
a^2\epsilon_{ij}x_j/2x_k x_k\right)^2/2\mu\to 0,$
the Hamiltonian $H_{\perp}$ reduces to
$H_0=\mu\omega_P^2 x_i x_i/2.$
In the limit of $\mu\to 0$ in the Lagrangian formalism, the
Lagrangian $L$ reduces to
$L_0=-\mu\omega_c \epsilon_{ij}\dot{x_i}x_j/2-\mu\omega_0
a^2\epsilon_{ij}\dot{x_i}x_j/2x_k x_k -\mu\omega_P^2 x_i x_i/2.$
The canonical momenta is
$p_{0i}=\partial L_0/\partial \dot{x_i}=-\mu\omega_c
\epsilon_{ij}x_j/2-\mu\omega_0 a^2\epsilon_{ij}x_j/2x_k x_k,$
and the corresponding Hamiltonian
$H_0^{\prime}=p_{0i}\dot{x_i}-L_0=\mu\omega_P^2 x_i x_i/2=H_0.$
Thus the limit of vanishing mechanical kinetic energy in the
Hamiltonian formalism identifies with the limit of the mass
$\mu\to 0$ in the Lagrangian formalism.
The massless limit was studied by Dunne, Jackiw and Trugenberger
\cite{DJT90}.
The first term of the equation (\ref{Eq:H2}) shows that in the
limit $\mu \to 0$ there are constraints \cite{Baxt,JZZ96}
\begin{equation}
\label{Eq:Ci}
\varphi_i=p_i +\frac{1}{2}\mu\omega_c
\epsilon_{ij}x_j+\frac{1}{2}\mu\omega_0
a^2\frac{\epsilon_{ij}x_j}{x_k x_k}=0,
\end{equation}
which should be carefully treated. The subject was treated simply
by the symplectic method in \cite{FJ,DJ93}. In this paper we work
in the approach of the Dirac brackets. The Poisson brackets of the
constraints (\ref{Eq:Ci}) are
\begin{equation}
\label{Eq:Poisson-1}
\{\varphi_i, \varphi_i\}_P= \mu\omega_c\epsilon_{ij}\ne 0,
\end{equation}
so that the corresponding Dirac brackets of $\{\varphi_i,
x_j\}_D$, $\{\varphi_i, p_j\}_D$, $\{x_i, x_j\}_D$,  $\{p_i,
p_j\}_D$ and $\{x_i, p_j\}_D$ can be determined. The Dirac
brackets of $\varphi_i$  with any variables $x_i$ and $p_j$ are
zero so that the constraints (\ref{Eq:Ci}) are strong conditions.
It can be used to eliminate dependent variables. If we select
$x_1$ and $x_2$ as the independent variables, from the constraints
(\ref{Eq:Ci}) the variables $p_1$ and $p_2$ can be represented by,
respectively, the independent variables $x_2$ and $x_1$. The Dirac
brackets of $x_1$ and $x_2$ is
\begin{equation}
\label{Eq:Dirac}
\{x_1,x_2\}_D=1/\mu\omega_c.
\end{equation}
Introducing new canonical variables $x=x_1$  and $p=\mu\omega_c
x_2,$ we have the corresponding commutation relation
$[x,p]=i\hbar$. Defining the effective mass
$\mu^{\ast}\equiv \mu\omega_c^2/\omega_P^2$
and the effective frequency
$\omega^{\ast} \equiv\omega_P^2/\omega_c,$
the Hamiltonian $H_0$ is represented as
$H_0=p^2/2\mu^{\ast}+\mu^{\ast}{\omega^{\ast}}^2x^2/2.$
Introducing an annihilation operator
$A= \sqrt{\mu^{\ast}\omega^{\ast}/2\hbar}\;q
+i\sqrt{1/2\hbar\mu^{\ast}\omega^{\ast}}\;p,$
we obtain
$H_0=\hbar\omega^{\ast}\left(A^\dagger A+1/2\right).$
The operators $A$ and $A^\dagger$ satisfies $[A,A^\dagger]=1$. The
eigenvalues of the number operator $N=A^\dagger A$ is $n=0, 1, 2,
\cdots$.

Now we consider the canonical angular momentum of an ion. Using
Eq.~(\ref{Eq:Ci}) to replace $p_1$ and $p_2$ by, respectively, the
independent variables $x_2$ and $x_1$, the canonical angular
momentum $J_z=\epsilon_{ij}x_i p_j$ is rewritten as
$J_z=q\Phi_0/2\pi c+\mu\omega_c(x_1^2+x_2^2)/2,$
where $\Phi_0=\pi a^2 B_0$ is the magnetic flux inside the
solenoid. Similarly, using $A$ and $A^\dagger$ to rewrite $J_z$,
we obtain
\begin{equation}
\label{Eq:J-1}
J_z=\frac{q}{2\pi c}\Phi_0+\hbar\left(A^\dagger
A+\frac{1}{2}\right).
\end{equation}
The zero-point angular momentum of $J_z$ is
$\mathcal{J}_0=1/2+q\Phi_0/2\pi c.$ In the above the term
\begin{equation}
\label{Eq:J-AB}
\mathcal{J}_{AB}=\frac{q}{2\pi c}\Phi_0
\end{equation}
is the zero-point angular momentum induced by the AB vector
potential. $\mathcal{J}_{AB}$ takes fractional values. It is
related to the region where the magnetic field ${\bf B}_{0,out}=0$
but the corresponding vector potential ${\bf A}_{out} \ne 0.$

It is worth noting that here ${\bf B}_c$, like a ``spectator",
does not contribute to $\mathcal{J}_{AB}$. In order to clarify the
role played by ${\bf B}_c$, we consider the case of ${\bf B}_c=0$.
In this case the modified combined trap is as stable as a Paul
trap. The corresponding mechanical kinetic energy reduces to
$\tilde E_k=\mu \dot{x_i} \dot{x_i}/2=\left(p_i+\mu\omega_0
a^2\epsilon_{ij}x_j/2x_k x_k\right)^2/2\mu$.
In the limit of $\tilde E_k\to 0$ the reduced constraints are
$\tilde \varphi_i=p_i+\mu\omega_0 a^2\epsilon_{ij}x_j/2x_k x_k=0.$
Here the special feature is that the corresponding Poisson
brackets
\begin{equation}
\label{Eq:Poisson-2}
\{\tilde{\varphi_i}, \tilde{\varphi_j}\}_P\equiv0,
\end{equation}
so that the Dirac brackets $\{\tilde{\varphi_i}, x_j\}_D$,
$\{\tilde{\varphi_i}, p_j\}_D$, $\{x_i, x_j\}_D$, $\{p_i, p_j\}_D$
and $\{x_i, p_j\}_D$ cannot be determined. There is no way to
establish dynamics at the quantum mechanics level. It means that
the AB vector potential alone cannot lead to non-trivial dynamics
survived in the limit of $\tilde E_k\to 0$ at the quantum
mechanics level. It is clear that though the ``spectator" magnetic
field ${\bf B}_c$ does not contribute to $\mathcal{J}_{AB}$, it
plays essential role in guaranteeing non-trivial dynamics survived
in the limit of vanishing mechanical kinetic energy at the quantum
mechanical level.

What is the essential difference between ${\bf A}_{out}$ and ${\bf
A}_c$ in the region of ${\bf B}_{0,out}=0?$ For this purpose we
consider the case of ${\bf B}_{0,in}=0.$ In this case the modified
combined trap reduces to a combined trap.  The corresponding
mechanical kinetic energy reduces to
$\hat E_k=\mu \dot{x_i} \dot{x_i}/2=\left(p_i+\mu\omega_c
\epsilon_{ij}x_j/2\right)^2/2\mu$.
In the limit of $\hat E_k\to 0$ the reduced constraints are
$\hat \varphi_i=p_i+\mu\omega_c \epsilon_{ij}x_j/2=0.$ The Poisson
brackets of the reduce constraints $\hat \varphi_i$ are the same
as ones of the constraints $\varphi_i$ in
Eq.~(\ref{Eq:Poisson-1}):
\begin{equation}
\label{Eq:Poisson-3}
\{\hat \varphi_i, \hat \varphi_i\}_P= \mu\omega_c\epsilon_{ij}\ne
0.
\end{equation}
Thus there is non-trivial dynamics survived in the limit of
vanishing mechanical kinetic energy at the quantum mechanical
level. By the similar procedure of treating the constraints
(\ref{Eq:Ci}), we obtain the angular momentum
$\hat J_z=\hbar\left(A^\dagger A+\frac{1}{2}\right)$
with the zero-point angular momentum
$\mathcal{\hat J}_0=1/2.$
This elucidates the essential difference between ${\bf A}_{out}$
and ${\bf A}_c$ in the region of ${\bf B}_{0,out}=0:$ The ${\bf
A}_c$ alone can lead to non-trivial dynamics survived in the limit
of vanishing mechanical kinetic energy at the quantum mechanical
level; But the ${\bf A}_{0,out}$ alone cannot.

As is well known, we can perform a gauge transformation $\chi$ so
that the resulting vector potential
${\bf A}_{out}^{\prime}={\bf A}_{out}+\nabla \chi =0.$
A suitable gauge function \cite{note-4} is
$\chi =-B_0 a^2 \phi(x_1,x_2)/2$
with the polar angle
$\phi(x_1,x_2)=tan^{-1}(x_2/x_1).$
%
In the Schr\"odinger equation the corresponding gauge
transformation is $\mathcal{G}=exp(iq\chi/c\hbar).$ Under this
gauge transformation the Hamiltonian $H_{\perp}(x_1,x_2)$ in
Eq.~(\ref{Eq:H2}) is transformed into
$H_{\perp} \to
\mathcal{G}H_{\perp}\mathcal{G}^{-1}=H_{\perp}^{\prime}
=\left(p_i+\mu\omega_c
\epsilon_{ij}x_j/2\right)^2/2\mu+\mu\omega_P^2 x_i^2/2.$
In the limit of vanishing mechanical kinetic energy the
corresponding reduced constraints are
$\hat\varphi_i=p_i+\mu\omega_c \epsilon_{ij}x_j/2=0.$
Under the gauge transformation $\mathcal{G}$ the angular momentum
$J_z$ is transformed into
$J_z \to
\mathcal{G}J_z\mathcal{G}^{-1}=J_z^{\prime}=x_1p_2-x_2p_1+
q\Phi_0/2\pi c.$
Using the above reduced constraints $\hat\varphi_i$ to represent
$p_1$ and $p_2$ by, respectively, the independent variables $x_2$
and $x_1$, the first term in $J_z^{\prime}$ reads
$x_1p_2-x_2p_1=\mu\omega_c(x_1^2+x_2^2)/2.$
Thus
$J_z^{\prime}=J_z=q\Phi_0/2\pi c+\mu\omega_c(x_1^2+x_2^2)/2$.
This result shows that the fractional
zero-point canonical angular momentum induced by the AB vector
potential is a real physical observable which cannot be gauged
away by a gauge transformation.

\vspace{0.4cm}

In summary, in this paper unexpected new features of the AB
effects are revealed. In a limit of vanishing mechanical kinetic
energy the AB vector potential alone cannot induce a fractional
zero-point canonical angular momentum on charged particles at the
quantum mechanical level in the AB magnetic field-free region. The
induced effects essentially depends upon the participation of a
``spectator" magnetic field. These results are significance for
both theories and experiments. Theoretically, the physical role
played by the AB vector potential is subtle which needs to
carefully analyze at the full quantum mechanical level.
Experimentally, real measurements of such an induced fractional
zero-point canonical angular momentum may need to involve
``spectator" magnetic fields.

\vspace{0.4cm}

This work has been supported by the Natural Science Foundation of
China under the grant number 10575037 and by the Shanghai
Education Development Foundation.


\begin{thebibliography}{99}
\bibitem{note-4}
This gauge function is singular at $x_1=0$. The values of the
polar angle $\phi$ on both sides of $x_2=0$, $x_1<0$ differ by
$2\pi$.  There is a cut along the negative semi-axis of $x_1$, the
so called Dirac string. However, there is no need to perform a
singular gauge transformation.
\end{thebibliography}
\end{document}